\documentclass[twocolumn,showpacs,preprintnumbers]{revtex4}
\usepackage{graphicx}
\usepackage{dcolumn}
\usepackage{bm}
\usepackage{amssymb}
\usepackage{amsmath}
\usepackage{graphicx}

\input{tcilatex}

\begin{document}

\title{Multipartite Entanglement Measure}
\author{Chang-shui Yu}
\email{quaninformation@sina.com}
\author{He-shan Song}
\affiliation{Department of Physics, Dalian University of Technology,\\
Dalian 116024, China.}
\date{\today}

\begin{abstract}
In this paper, we generalize the residual entanglement to the case of
multipartite states in arbitrary dimensions by making use of a new method.
Through the introduction of a special entanglement measure, the residual
entanglement of mixed states takes on a form that is more elegant than that
in Ref.[7] (Phys.Rev.A \textbf{61 }(2000) 052306) . The result obtained in
this paper is different from the previous one given in Ref.[8] (Phys.Rev.A 
\textbf{63 }(2000) 044301). Several examples demonstrate that our present
result is a good measurement of the multipartite entanglement. Furthermore,
the original residual entanglement is a special case of our result.
\end{abstract}

\maketitle



\section{\protect\bigskip Introduction}

Entanglement is the interesting feature which distinguishes the quantum
world from the classical one. It has become a useful physical resource in
quantum information processing (QIP) that has undergone a rapid development
in recent years [1,2]. Because quantum entangled states play an important
role in strorage and transport of quantum information, a great attention has
been focused on the study of their properites. Entanglement has been
quantified through the concept of concurrence that was introduced by
Wootters in 1998 [3]. However, Wootters' concurrence is available only for
bipartite systems with two levels, which, as known to all, is developed to
different extents later [4,5,6]. The quantification of multipartite
entanglement (multi-way entanglement) is still an open problem.

In 2000, Valerie Coffman, Joydip Kundu and William K. Wootters [7] shed new
light on the understanding of multipartite entanglement by the study of
distributed entanglement. They discovered an interesting quantity for a
tripartite two-level system, referred to as the residual entanglement that
is defined by%
\begin{equation}
\tau _{ABC}=C_{A(BC)}^{2}-C_{AB}^{2}-C_{AC}^{2},
\end{equation}%
where $C_{AB}$ and $C_{AC}$ are the concurrences of the original pure state $%
\rho _{ABC}$ with traces taken over qubits $C$ and $B$, respectively. $%
C_{A(BC)}$ is the concurrence of $\rho _{A(BC)}$ with qubits $B$ and $C$
regarded as a single object. In Ref.[7], it is shown that the residual
entanglement of the pure two-level state $\left\vert \psi \right\rangle =%
\underset{ijk}{\sum }a_{ijk}\left\vert ijk\right\rangle $ is given by%
\begin{equation}
\tau _{ABC}=2\left\vert \sum a_{ijk}a_{i^{\prime }j^{\prime
}m}a_{npk}a_{n^{\prime }p^{\prime }m^{\prime }}\epsilon _{ii^{\prime
}}\epsilon _{jj^{\prime }}\epsilon _{mm^{\prime }}\epsilon _{nn^{\prime
}}\epsilon _{pp^{\prime }}\right\vert ,
\end{equation}%
where the summation over repeated indices is implied and $\epsilon _{\alpha
\beta }=-\epsilon _{\beta \alpha }=\delta _{\alpha \beta }$. Since the
residual entanglement is unchanged by permutations of $A$, $B$ and $C$, it
can be regarded as representing a collective property of three qubit and can
be used to quantify the 3-way entanglement [8].

In 2001, Alexander Wong and Nelson Christensen [8] demonstrated a
generalization of the 3-tangle for $n$ qubits. Starting from the definition
of the $n$-tangle in a form similar to eq.(2), that is a generalization of
the concurrence of pure states with three or an even number of qubits, they
obtained a result with elegant form analogous to the result in Ref.[7].
Their result has also the property that $\tau _{1\cdots n}$ is unchanged by
permutations. However, $\tau _{1\cdots n}$ itself is not a measure of n-way
entanglement.

In this paper, starting from eq.(1) and the generalized bipartite
entanglement measure in arbitrary dimensions [5,6,10], we generalize the
residual entanglement to multipartite systems in arbitrary dimensions in an
approach similar to Ref.[7]. Our result provides a good measure for n-way
entanglement in arbitrary dimensions. As to the case discussed in Ref.[7],
the residual entanglement given in this paper can be easily reduced to the
original one. The paper is organized as follows: Starting with the
entanglement measure for the bipartite pure state-linear entropy, we
generalize it to the case of mixed state; then we prove that the inequality
similar to $C_{A(BC)}^{2}\geqslant C_{AB}^{2}+C_{AC}^{2}$ mentioned in
Ref.[7] holds for tripartite pure states in arbitrary dimensions, generalize
it to the case of mixed state and define the residual entanglement something
different from the original one; and then we generalize the residual
entanglement to the case of multipartite systems; finally we give several
examples to demonstrate our generalization can work well for the
multipartite entanglement in its right.

\section{The Special Entanglement Measure For Bipartite Systems}

The linear entropy of a pure state $\varphi $ is defined by%
\begin{equation*}
E(\varphi )=1-Tr(\rho _{\alpha }^{2}),
\end{equation*}%
where $\rho _{\alpha }$ denotes the reduced density matrix of a bipartite
system. The pure state concurrence in arbitrary dimensions [9] is defined by 
\begin{equation*}
C(\psi )=\sqrt{2(1-Tr(\rho _{\alpha }^{2}))},
\end{equation*}%
which is the same to the concurrence [10], i.e.%
\begin{equation}
C=|\mathbf{C}|=\sqrt{\underset{\alpha =1}{\overset{N_{1}(N_{1}-1)/2}{\sum }%
\text{ }}\overset{N_{2}(N_{2}-1)/2}{\underset{\beta =1}{\sum }}\left\vert
C_{\alpha \beta }\right\vert ^{2}},
\end{equation}%
where C$_{\alpha \beta }$ =$\left\langle \psi |\widetilde{\psi }_{\alpha
\beta }\right\rangle $, $\left\vert \widetilde{\psi }_{\alpha \beta
}\right\rangle =\left( L_{\alpha }\otimes L_{\beta }\right) \left\vert \psi
^{\ast }\right\rangle $ with $L_{\alpha }$, \ $L_{\beta }$ denoting
generators of $SO(N_{1})$ and $SO(N_{2})$ respectively. Therefore, 
\begin{equation}
2E(\varphi )=C^{2}=\underset{\alpha =1}{\overset{N_{1}(N_{1}-1)/2}{\sum }%
\text{ }}\overset{N_{2}(N_{2}-1)/2}{\underset{\beta =1}{\sum }}\left\vert
C_{\alpha \beta }\right\vert ^{2}.
\end{equation}

If we employ the linear entropy $C^{\prime }(\varphi )=2E(\varphi )$ as
entanglement measure for pure states, we can generalize it to the case of
mixed states.

The mixed states $\rho =\sum\limits_{k}\omega _{k}\left\vert \psi
^{k}\right\rangle \left\langle \psi ^{k}\right\vert $ can be written in
matrix notation as $\rho =\Psi W\Psi ^{\dagger }$, where $W$ is a diagonal
matrix with $W_{kk}=\omega _{k}$, the columns of \ the matrix $\Psi $
correspond to the vectors $\psi ^{k}$. Consider the eigenvalue
decomposition, $\rho =\Phi M\Phi ^{\dagger }$, where $M$ is a diagonal
matrix whose diagonal elements are the eigenvalues of $\rho $, and $\Phi $
is a unitary matrix whose columns are the eigenvectors of $\rho $. From
Ref.[5], one can get $\Psi W^{1/2}=\Phi M^{1/2}T$, where $T$ is a
Right-unitary matrix. The mixed states are separable iff there exist a
decomposition such that $\psi ^{k}$ for every $k$ is separable. The
entanglement measure of formation can be defined as the infimum of the
average $\left\vert C^{\prime }(\psi ^{k})\right\vert $. Namely, $C(\rho
)=\inf \sum\limits_{k}\omega _{k}\left\vert C^{\prime }(\psi
^{k})\right\vert $, if $C(\rho )$ is assigned as the entanglement measure
for tripartite mixed states. Therefore, for any a decomposition 
\begin{equation*}
\rho =\sum\limits_{k}\omega _{k}\left\vert \psi ^{k}\right\rangle
\left\langle \psi ^{k}\right\vert ,
\end{equation*}%
considering eq.(4), 
\begin{equation*}
\sum\limits_{k}\omega _{k}C^{\prime }(\psi ^{k})=2\sum\limits_{k}\omega
_{k}E(\psi ^{k})
\end{equation*}%
\begin{equation}
=\sum\limits_{k}\omega _{k}\left( \underset{\alpha =1}{\overset{%
N_{1}(N_{1}-1)/2}{\sum }\text{ }}\overset{N_{2}(N_{2}-1)/2}{\underset{\beta
=1}{\sum }}\left\vert C_{\alpha \beta }^{k}\right\vert ^{2}\right) .
\end{equation}%
According to the convexity property of eq.(4), we have,%
\begin{equation*}
\sum\limits_{k}\omega _{k}C^{\prime }(\psi ^{k})\geqslant \underset{\alpha =1%
}{\overset{N_{1}(N_{1}-1)/2}{\sum }\text{ }}\overset{N_{2}(N_{2}-1)/2}{%
\underset{\beta =1}{\sum }}\left( \sum\limits_{k}\omega _{k}\left\vert
\left\langle \psi ^{k}|\widetilde{\psi }_{\alpha \beta }^{k}\right\rangle
\right\vert \right) ^{2}
\end{equation*}%
\begin{equation}
=\underset{\alpha =1}{\overset{N_{1}(N_{1}-1)/2}{\sum }\text{ }}\overset{%
N_{2}(N_{2}-1)/2}{\underset{\beta =1}{\sum }}\left( \sum\limits_{k}\omega
_{k}\left\vert \left\langle \psi ^{k}\right\vert S_{\alpha \beta }\left\vert
\left( \psi ^{k}\right) ^{\ast }\right\rangle \right\vert \right) ^{2}
\end{equation}%
with $S_{\alpha \beta }=L_{\alpha }\otimes L_{\beta }$. If the inequality $%
\sum\limits_{i}|x_{i}|\geqslant \left\vert \sum\limits_{i}x_{i}\right\vert $
and the Cauchy-Schwarz inequality 
\begin{equation}
\left( \sum\limits_{i}x_{i}^{2}\right) ^{1/2}\left(
\sum\limits_{i}y_{i}^{2}\right) ^{1/2}\geqslant \sum\limits_{i}x_{i}y_{i},
\end{equation}%
are considered, eq.(6) can be written as 
\begin{eqnarray*}
&&\sum\limits_{k}\omega _{k}C^{\prime }(\psi ^{k}) \\
&\geqslant &\underset{\alpha =1}{\overset{N_{1}(N_{1}-1)/2}{\sum }\text{ }}%
\overset{N_{2}(N_{2}-1)/2}{\underset{\beta =1}{\sum }}\left(
\sum\limits_{k}\left\vert T^{\dag }M^{1/2}\Phi ^{\dag }S_{\alpha \beta }\Phi
^{\ast }M^{1/2}T^{\ast }\right\vert {}_{kk}\right) ^{2}
\end{eqnarray*}%
\begin{equation*}
=\underset{\alpha =1}{\overset{N_{1}(N_{1}-1)/2}{\sum }\text{ }}\overset{%
N_{2}(N_{2}-1)/2}{\underset{\beta =1}{\sum }}\left(
\sum\limits_{k}\left\vert (T^{T}A_{\alpha \beta }T)\right\vert _{kk}\right)
^{2}
\end{equation*}%
\begin{equation}
\geqslant \left[ \sum\limits_{k}\left\vert T^{T}\left( \underset{\alpha =1}{%
\overset{N_{1}(N_{1}-1)/2}{\sum }\text{ }}\overset{N_{2}(N_{2}-1)/2}{%
\underset{\beta =1}{\sum }}z_{\alpha \beta }A_{\alpha \beta }\right)
T\right\vert _{kk}\right] ^{2}
\end{equation}%
in matrix notation of $\rho $, where $A_{\alpha \beta }=M^{1/2}\Phi
^{T}S_{\alpha \beta }\Phi M^{1/2}$ and $z_{\alpha \beta }=y_{\alpha \beta
}e^{i\phi }$ with $y_{\alpha \beta }>0$, $\sum\limits_{\alpha \beta
}y_{\alpha \beta }^{2}=1$. By virtue of the definition of meaure of
entanglement of formation, one can get%
\begin{equation}
C^{\prime }(\rho )=\inf_{T}\left[ \sum\limits_{k}\left\vert T^{T}\left( 
\underset{\alpha =1}{\overset{N_{1}(N_{1}-1)/2}{\sum }\text{ }}\overset{%
N_{2}(N_{2}-1)/2}{\underset{\beta =1}{\sum }}z_{\alpha \beta }A_{\alpha
\beta }\right) T\right\vert _{kk}\right] ^{2}
\end{equation}%
for mixed states. The infimum is given by $\left[ \underset{z\in C^{\alpha
\beta }}{max}\lambda _{1}(z)-\sum\limits_{i>1}\lambda _{i}(z)\right] ^{2}$
with $\lambda _{j}(z)$ are the singular values of $\underset{\alpha =1}{%
\overset{N_{1}(N_{1}-1)/2}{\sum }\text{ }}\overset{N_{2}(N_{2}-1)/2}{%
\underset{\beta =1}{\sum }}z_{\alpha \beta }A_{\alpha \beta }$. Hence we get
the lower bound%
\begin{equation}
C^{\prime }(\rho )=\left[ \underset{z\in C^{\alpha \beta }}{max}\lambda
_{1}(z)-\sum\limits_{i>1}\lambda _{i}(z)\right] ^{2}=\left[ C(\rho )\right]
^{2},
\end{equation}%
which is consistent to the result in Ref.[5,6] neglecting the square.

\section{Residual Entanglement of Tripartite Systems}

Consider a given tripartite pure state $\left\vert \psi \right\rangle _{ABC}$
in arbitrary dimension. For the subsystem $\rho _{AB}$ made up of $A$ and $B$
by tracing over $C$, similar to ref.[7], we can employ the eq.(10) to write
the following inequality%
\begin{equation*}
C_{AB}^{\prime }(\rho )=\left[ \underset{z\in C^{\alpha \beta }}{max}\lambda
_{1}(z)-\sum\limits_{i>1}\lambda _{i}(z)\right] ^{2}\leq
Tr(Q_{AB}Q_{AB}^{\dag })
\end{equation*}%
\begin{eqnarray*}
&\leq &Tr(\underset{\alpha =1}{\overset{N_{1}(N_{1}-1)/2}{\sum }\text{ }}%
\overset{N_{2}(N_{2}-1)/2}{\underset{\beta =1}{\sum }}\left\vert z_{\alpha
\beta }\right\vert \left\vert A_{\alpha \beta }\right\vert )^{2} \\
&\leq &Tr[\left( \underset{\alpha =1}{\overset{N_{1}(N_{1}-1)/2}{\sum }\text{
}}\overset{N_{2}(N_{2}-1)/2}{\underset{\beta =1}{\sum }}\left\vert z_{\alpha
\beta }\right\vert ^{2}\right) ^{1/2} \\
&&\times \left( \underset{\alpha =1}{\overset{N_{1}(N_{1}-1)/2}{\sum }\text{ 
}}\overset{N_{2}(N_{2}-1)/2}{\underset{\beta =1}{\sum }}\left\vert A_{\alpha
\beta }\right\vert ^{2}\right) ^{1/2}]^{2} \\
&=&\underset{\alpha =1}{\overset{N_{1}(N_{1}-1)/2}{\sum }\text{ }}\overset{%
N_{2}(N_{2}-1)/2}{\underset{\beta =1}{\sum }}Tr\left\vert M^{1/2}\Phi ^{\dag
}S_{\alpha \beta }\Phi ^{\ast }M^{1/2}\right\vert ^{2} \\
&=&\underset{\alpha =1}{\overset{N_{1}(N_{1}-1)/2}{\sum }\text{ }}\overset{%
N_{2}(N_{2}-1)/2}{\underset{\beta =1}{\sum }}Tr\left( \rho _{AB}\left( 
\widetilde{\rho }_{AB}\right) _{\alpha \beta }\right) ,
\end{eqnarray*}%
i.e.%
\begin{equation}
C_{AB}^{\prime }(\rho )\leq \underset{\alpha =1}{\overset{N_{1}(N_{1}-1)/2}{%
\sum }\text{ }}\overset{N_{2}(N_{2}-1)/2}{\underset{\beta =1}{\sum }}%
Tr\left( \rho _{AB}\left( \widetilde{\rho }_{AB}\right) _{\alpha \beta
}\right) ,
\end{equation}%
where $Q_{AB}=$ $\underset{\alpha =1}{\overset{N_{1}(N_{1}-1)/2}{\sum }\text{
}}\overset{N_{2}(N_{2}-1)/2}{\underset{\beta =1}{\sum }}z_{\alpha \beta
}A_{\alpha \beta }$, $\left( \widetilde{\rho }_{AB}\right) _{\alpha \beta
}=S_{\alpha \beta }\rho _{AB}^{\ast }S_{\alpha \beta }$ and the other
parameters are defined the same as above section. For the subsystem $\rho
_{AC}$, the inequality analogous to (11) holds. Therefore, we can bound the
sum $C_{AB}^{\prime }(\rho )+C_{AC}^{\prime }(\rho ):$%
\begin{eqnarray}
&&C_{AB}^{\prime }(\rho )+C_{AC}^{\prime }(\rho )  \notag \\
&\leq &\underset{\alpha =1}{\overset{N_{1}(N_{1}-1)/2}{\sum }\text{ }}%
\overset{N_{2}(N_{2}-1)/2}{\underset{\beta =1}{\sum }}Tr\left( \rho
_{AB}\left( \widetilde{\rho }_{AB}\right) _{\alpha \beta }\right)  \notag \\
&&+\underset{\gamma =1}{\overset{N_{1}(N_{1}-1)/2}{\sum }\text{ }}\overset{%
N_{2}^{\prime }(N_{2}^{\prime }-1)/2}{\underset{\delta =1}{\sum }}Tr\left(
\rho _{AC}\left( \widetilde{\rho }_{AC}\right) _{\gamma \delta }\right) .
\end{eqnarray}%
If we write the pure state $\left\vert \psi \right\rangle _{ABC}$ in the
standard basis defined in $n_{1}\times n_{2}\times n_{3}$ dimension, i.e. 
\begin{equation*}
\left\vert \psi \right\rangle _{ABC}=\underset{ijk}{\sum }a_{ijk}\left\vert
ijk\right\rangle ,
\end{equation*}%
where $i=0,1,\cdots n_{1}-1$, $j=0,1,\cdots n_{2}-1$ and $k=0,1,\cdots
n_{3}-1$. Considering the coefficients $a_{ijk}$, one can obtain the
following three equations 
\begin{eqnarray}
&&\underset{\alpha =1}{\overset{N_{1}(N_{1}-1)/2}{\sum }\text{ }}\overset{%
N_{2}(N_{2}-1)/2}{\underset{\beta =1}{\sum }}Tr\left( \rho _{AB}\left( 
\widetilde{\rho }_{AB}\right) _{\alpha \beta }\right)  \notag \\
&=&\sum a_{ijk}a_{mnk}^{\ast }\epsilon _{mm^{\prime }}\epsilon _{nn^{\prime
}}a_{m^{\prime }n^{\prime }p}^{\ast }a_{i^{\prime }j^{\prime }p}\epsilon
_{i^{\prime }i}\epsilon _{j^{\prime }j},
\end{eqnarray}

\begin{eqnarray}
&&\overset{N_{1}(N_{1}-1)/2}{\underset{\gamma =1}{\sum }}\overset{%
N_{2}^{\prime }(N_{2}^{\prime }-1)/2}{\underset{\delta =1}{\sum }}Tr\left(
\rho _{AC}\left( \widetilde{\rho }_{AC}\right) _{\alpha \beta }\right) 
\notag \\
&=&\sum a_{ikj}a_{mkn}^{\ast }\epsilon _{mm^{\prime }}\epsilon _{nn^{\prime
}}a_{m^{\prime }pn^{\prime }}^{\ast }a_{i^{\prime }pj^{\prime }}\epsilon
_{i^{\prime }i}\epsilon _{j^{\prime }j},
\end{eqnarray}%
\begin{eqnarray*}
&&\underset{\alpha =1}{\overset{N_{1}(N_{1}-1)/2}{\sum }\text{ }}\overset{%
N_{2}(N_{2}-1)/2}{\underset{\beta =1}{\sum }}\left\vert \left\langle \psi
^{k}|\widetilde{\psi }_{\alpha \beta }^{k}\right\rangle \right\vert ^{2} \\
&=&\underset{\alpha =1}{\overset{N_{1}(N_{1}-1)/2}{\sum }\text{ }}\overset{%
N_{2}(N_{2}-1)/2}{\underset{\beta =1}{\sum }}\left\langle \psi ^{k}|%
\widetilde{\psi }_{\alpha \beta }^{k}\right\rangle \left\langle \widetilde{%
\psi }_{\alpha \beta }^{k}|\psi ^{k}\right\rangle \\
&=&\underset{\alpha =1}{\overset{N_{1}(N_{1}-1)/2}{\sum }\text{ }}\overset{%
N_{2}(N_{2}-1)/2}{\underset{\beta =1}{\sum }}Tr\left( \rho _{A\left(
BC\right) }\widetilde{\rho }_{A\left( BC\right) }\right)
\end{eqnarray*}%
\begin{equation}
=\sum a_{ijk}a_{mnk^{\prime }}^{\ast }\epsilon _{mm^{\prime }}\epsilon
_{\left( nk^{\prime }\right) ,\left( n^{\prime }p\right) }a_{m^{\prime
}n^{\prime }p}^{\ast }a_{i^{\prime }j^{\prime }p^{\prime }}\epsilon
_{i^{\prime }i}\epsilon _{\left( j^{\prime }p^{\prime }\right) \left(
jk\right) }.
\end{equation}%
where 
\begin{equation*}
\epsilon _{mm^{\prime }}\epsilon _{ii^{\prime }}=\delta _{mi}\delta
_{m^{\prime }i^{\prime }}-\delta _{mi^{\prime }}\delta _{m^{\prime }i},
\end{equation*}%
and 
\begin{equation*}
\epsilon _{\left( nk^{\prime }\right) ,\left( n^{\prime }p\right) }\epsilon
_{\left( j^{\prime }p^{\prime }\right) \left( jk\right) }=\delta _{\left(
nk^{\prime }\right) ,\left( j^{\prime }p^{\prime }\right) }\delta _{\left(
n^{\prime }p\right) ,\left( jk\right) }-\delta _{\left( nk^{\prime }\right)
,\left( jk\right) }\delta _{\left( n^{\prime }p\right) ,\left( j^{\prime
}p^{\prime }\right) }.
\end{equation*}%
\ After simplification of eq.(13), eq.(14) and eq.(15), one can find
that\bigskip 
\begin{eqnarray}
&&\underset{\alpha =1}{\overset{N_{1}(N_{1}-1)/2}{\sum }\text{ }}\overset{%
N_{2}(N_{2}-1)/2}{\underset{\beta =1}{\sum }}Tr\left( \rho _{AB}\left( 
\widetilde{\rho }_{AB}\right) _{\alpha \beta }\right)  \notag \\
&&+\underset{\gamma =1}{\overset{N_{1}(N_{1}-1)/2}{\sum }\text{ }}\overset{%
N_{2}^{\prime }(N_{2}^{\prime }-1)/2}{\underset{\delta =1}{\sum }}Tr\left(
\rho _{AC}\left( \widetilde{\rho }_{AC}\right) _{\alpha \beta }\right) 
\notag \\
&=&\underset{\alpha =1}{\overset{N_{1}(N_{1}-1)/2}{\sum }\text{ }}\overset{%
N_{2}(N_{2}-1)/2}{\underset{\beta =1}{\sum }}\left\vert \left\langle \psi
^{k}|\widetilde{\psi }_{\alpha \beta }^{k}\right\rangle \right\vert ^{2}.
\end{eqnarray}%
Therefore, we have%
\begin{equation}
C_{AB}^{\prime }(\rho )+C_{AC}^{\prime }(\rho )\leq C_{A(BC)}^{\prime }(\rho
).
\end{equation}%
According to the result given in the above section, (17) can\ also be
written as%
\begin{equation}
C_{AB}^{2}(\rho )+C_{AC}^{2}(\rho )\leq C_{A(BC)}^{2}(\rho ).
\end{equation}

For mixed states, analogous to Ref.[7], we can also generalize the above
result. If there exists a mixed state $\rho _{ABC}=\sum p_{i}\left\vert \psi
_{i}\right\rangle \left\langle \psi _{i}\right\vert $, which is the
decomposition corresponding to the minimal $\sum p_{i}C^{\prime }(\left\vert
\psi _{i}\right\rangle )$, for every $\left\vert \psi _{i}\right\rangle $,
the inequality similar to (17) holds. Summarizing all the inequalies, we have%
\begin{equation}
\sum p_{i}C_{AB}^{\prime }(\rho _{i})+\sum p_{i}C_{AC}^{\prime }(\rho
_{i})\leq \sum p_{i}C_{A(BC)}^{\prime }(\rho _{i}).
\end{equation}%
According to the convex $C_{\alpha }^{2}(\rho )$, the left-hand side of the
inequality (19) is greater than or equal to $C_{AB}^{2}(\rho
)+C_{AC}^{2}(\rho )$; according to the definition of entanglement measure of
mixed bipartite states, the right-hand of the inequality is equal to $%
C_{A(BC)}^{\prime }(\rho )$. Therefore, we have 
\begin{equation*}
C_{AB}^{2}(\rho )+C_{AC}^{2}(\rho )\leq C_{A(BC)}^{\prime }(\rho
)=C_{A(BC)}^{2}(\rho )
\end{equation*}%
holds for the mixed state $\rho $.

If particle $A$ in $C_{A(BC)}^{2}$ is called the focus analogous to Ref.[7],
we find that the inequality holds independent on the choice of the focus. If
we introduce a quantity $\tau $ for every analogous inequality corresponding
every possible focus, then every one of the inequalities can be converted to
an equation. Namely, we can obtain%
\begin{equation*}
\tau _{A(BC)}+C_{AB}^{2}(\rho )+C_{AC}^{2}(\rho )=C_{A(BC)}^{2}(\rho ).
\end{equation*}%
\begin{equation*}
\tau _{B(AC)}+C_{AB}^{2}(\rho )+C_{BC}^{2}(\rho )=C_{B(AC)}^{2}(\rho ).
\end{equation*}%
\begin{equation*}
\tau _{C(AB)}+C_{CB}^{2}(\rho )+C_{AC}^{2}(\rho )=C_{C(AB)}^{2}(\rho ).
\end{equation*}%
Unlike the case of pure states in Ref.[7], it is difficult to tell whether $%
\tau _{A(BC)}$, $\tau _{B(AC)}$ and $\tau _{C(AB)}$ are equal because of the
introduction of $z_{\alpha \beta }$ in optimization, even for the pure state 
$\rho $ in arbitrary dimension. However, in Ref.[7], for tripartite states
with two levels, $\tau $ embodies a kind of global property, which can
measure 3-way entanglement, therefore, no matter whether $\tau _{\alpha }$s
are equal or not, it is reasonable to believe that $\tau _{\alpha }$s, with $%
\alpha $ corresponding to the different foci $A(BC)$, $B(AC)$ or $C(AB)$,
include a common quantity which embodies a kind of collective preperty,
independent of permutations and can be used to measure 3-way entanglement
just like $\tau $. What's more, we know that entanglement measure is
relative, hence we can define the minmal $\tau _{\alpha }$ as the residual
entanglement, which measures 3-way entanglement.

\section{Residual Entanglement of Multipartite Systems}

For an N-partite quantum state $\rho _{AB\cdots N}$ in arbitrary dimension,
one always regards it as a tripartite quantum state which can be assumed to
be $\rho _{AB\left( C\cdots N\right) }$, therefore the analogous inequality
holds%
\begin{equation}
C_{AB}^{2}(\rho )+C_{A\left( C\cdots N\right) }^{2}(\rho )\leq C_{A(BC\cdots
N)}^{2}(\rho ).
\end{equation}%
In the same way, one can get 
\begin{equation}
C_{AC}^{2}(\rho )+C_{A\left( D\cdots N\right) }^{2}(\rho )\leq C_{A\left(
C\cdots N\right) }^{2}(\rho ).
\end{equation}%
This iteration of the above inequalities leads to%
\begin{equation}
C_{AB}^{2}(\rho )+C_{AC}^{2}(\rho )+\cdots +C_{AN}^{2}(\rho )\leq
C_{A(BC\cdots N)}^{2}(\rho ).
\end{equation}%
Analogous to the last section, we have%
\begin{equation}
\tau _{A(BC\cdots N)}+C_{AB}^{2}(\rho )+C_{AC}^{2}(\rho )+\cdots
+C_{AN}^{2}(\rho )=C_{A(BC\cdots N)}^{2}(\rho ).
\end{equation}%
If changing the focus, one will obtain the other $N-1$ analogous equations.
It is worth noting that $AB$, $ABC$ and so on can all be regarded as an
object and all can be used as the focus. Therefore, there should exist $%
\underset{i=1}{\overset{\left[ \frac{N}{2}\right] }{\sum }}C_{N}^{i}$
analogous equations with $C_{N}^{i}=\frac{N!}{(N-i)!i!}$ and $\sum_{j}\in
\lbrack 1,\left[ \frac{N}{2}\right] ]$ in all, where $\left[ \frac{N}{2}%
\right] $ =$\left\{ 
\begin{array}{cc}
N/2, & N\text{ \ is even} \\ 
(N-1)/2, & N\text{ \ is odd}%
\end{array}%
\right. $. At the same time, there exist $\underset{i=1}{\overset{\left[ 
\frac{N}{2}\right] }{\sum }}C_{N}^{i}$ $\tau _{\alpha }$s which include a
common quantity that embodies the collective property of the given quantum
state, and independent on the choice of the focus (or permutations).
Therefore, analogous to the case of tripartite systems, we can also select
the minimal $\tau _{\alpha }$, where $\alpha $ belongs to the set of all the
different foci, as the residual entanglement. Now, the residual entanglement
can be written in a general form considering the tripartite case, in a more
rigorous way.

\textbf{Definition}.-The residual entanglement $\tau _{ABC\cdots N}$ of an $%
N $-particle system $\rho ^{ABC\cdots N}$ is defined as 
\begin{equation*}
\tau _{ABC\cdots N}=\min \{\tau _{\alpha }|\alpha =1,2,\cdots \underset{i=1}{%
,\overset{\left[ \frac{N}{2}\right] }{\sum }}C_{N}^{i}\},
\end{equation*}%
where $\alpha $ corresponds to all the possible foci.

\section{\protect\bigskip Examples}

For the generalized GHZ states [11] 
\begin{equation*}
\left\vert \psi \right\rangle =\frac{1}{\sqrt{2}}\left( \left\vert \underset{%
n}{\underbrace{0\cdots 0}}\right\rangle +\left\vert \underset{n}{\underbrace{%
1\cdots 1}}\right\rangle \right) ,
\end{equation*}%
no matter which particles are selected as the focus, the concurrence of the
state $\left\vert \psi \right\rangle $ which is regarded as a bipartite
state corresponding to the focus and the others, is $one$; and the
conccurrence of any subsystem by tracing over other $N-focus$ particles are $%
zero$. Hence, $\tau _{ABC\cdots N}=1$.

For the generalized $n$-qubit state 
\begin{eqnarray*}
\left\vert \phi \right\rangle &=&\alpha _{1}\left\vert 10\cdots
0\right\rangle +\alpha _{2}\left\vert 010\cdots 0\right\rangle +\alpha
_{3}\left\vert 0010\cdots 0\right\rangle \\
&&+\cdots +\alpha _{n}\left\vert 0000\cdots 1\right\rangle ,
\end{eqnarray*}%
our definition is reduced to the case introduced in Ref.[7], therefore, 
\begin{equation*}
C_{12}^{2}+C_{13}^{2}+\cdots +C_{1n}^{2}=C_{1\left( 2\cdots n\right) }^{2},
\end{equation*}%
namely, $\tau _{12\cdots n}=0$.

For N-particle product states, $\rho =\sum_{i}\rho _{1}^{i}\otimes \rho
_{2}^{i}$, where $\rho _{1}$ corresponds to $N_{1}$ particles and $\rho _{2}$
corresponds to the others, if the particles corresponding to $\rho _{1}$ (or 
$\rho _{2}$) are selected as the focus, we can obtain that the concurrence
of the two product subsystem $\rho _{1}$ and $\rho _{2}$ is zero, i.e. $%
C_{N_{1}N_{2}}^{2}(\rho )=0$. Since $\tau \geqslant 0$ is the minimum of all
the $\tau _{\alpha }$s, then $\tau =0$ in this case. Especially for the
four-particle pure product state $\Psi _{ABCD}$ of two singlet states
metioned in Ref.[8], one can obtain $\tau _{ABCD}=0$, which shows that the
state $\Psi _{ABCD}$ does not include 4-way entanglement. The result is
consistent to the fact and what was implied in Ref.[8].

\section{Conclusion and Discussion}

In summary, we introduced a special entanglement measure for bipartite
states. We generalize residual entanglement to the case of multipartite
states in arbitrary dimensions. Unlike Ref.[8], we generalize the residual
entanglement to the case of any $N-$partite system, and the examples show
that our residual entanglement can well measure the $n$-way entanglement.
Recalling the original residual entanglement [7], one can find that it is
just the special case of ours: it is a tripartite state with two levels (in
the case, we can get the same result). However, the special case easily
shows that $\tau _{A(BC)}$, $\tau _{B(AC)}$ and $\tau _{C(AB)}$ are equal
which have a more elegant form, but it is difficult to tell whether the same
relation holds in other cases. For tripartite states, it is impossible to
select two particles as a focus, there exist only three ways to select the
focus which is enough; but for multipartite states, $\tau _{\alpha }$
corresponding to only one particle as the focus is not enough to completely
embody the collective property, the other cases are valid and necessary.
Furthermore, one can find that the introduction of the special bipartite
entanglement measure in section II is just a medium of generalization of
residual entanglement from the pure states to the mixed states, which leads
to the difference between $C_{A(BC)}^{2}(\rho )$ in this paper and $\left(
C^{2}\right) _{A(BC)}^{\min }(\rho )$ in Ref.[7] for mixed states and
provides inestimable conveniences to generalize the residual entanglement
from tripartite case to multipartite one.

\section{Acknowledgement}

We thank X. X. Yi for extensive and valuable advice. We are grateful to
Fu-xiang Han for his revise. This work was supported by Ministry of Science
and Technology, China, under grant No.2100CCA00700.

\end{document}